\documentclass[12pt]{article}
\usepackage[english,portuguese]{babel}
\usepackage[utf8]{inputenc}
\usepackage{geometry}
\usepackage{authblk}
\usepackage{yfonts}
\usepackage{xcolor}
\usepackage{xfrac}
\usepackage{array}
\usepackage[c]{esvect}
\usepackage{setspace}
\usepackage{graphicx}
\usepackage{indentfirst} 
\usepackage{amsmath,amsfonts,amsthm,amscd,upref,amstext}
\usepackage{mathtools}
\usepackage{comment}
\usepackage{multirow}
\usepackage{url}
\usepackage[titletoc]{appendix}
\newcommand*{\rom}[1]{\expandafter\@slowromancap\romannumeral #1@}

\newtheorem{theorem}{Theorem}[section]

\newtheorem{defn}[theorem]{Definition}
\newtheorem{rmk}[theorem]{Remark}

\newcommand{\R}{\mathbb{R}}

\newcommand{\QLM}{QLM}
\newcommand{\QLE}{QLE}

\newcommand{\const}{const}

\newcommand{\pst}{\mathcal{M}^{3,1}}

\title{A proposal of quasi-local mass for $2$-surfaces of timelike mean curvature}
\author[1]{Bowen Zhao}
\author[2]{Shing-Tung Yau}
\author[1]{Lars Andersson}
\affil[1]{Beijing Institute of Mathematical Sciences and Applications, Beijing, China}
\affil[2]{Yau Mathmatical Sciences Center, Tsinghua University, Beijing, China}
\date{}                     
\begin{document}
\maketitle
\doublespacing

\selectlanguage{english}
\abstract{
A quasi-local mass, typically defined as an integral over a spacelike $2$-surface $\Sigma$, should encode information about the gravitational field within a finite, extended region bounded by $\Sigma$. Therefore, in attempts to quantize gravity, one may consider an infinite dimensional space of $2$-surfaces instead of an infinite dimensional space of $4$-dimensional Lorentzian spacetimes. However, existing definitions for quasilocal mass only applies to surfaces outside an horizon whose mean curvature vector is spacelike.
In this paper, we propose an extension of the Wang-Yau quasi-local energy/mass to surfaces with timelike mean curvature vector, including in particular trapped surfaces. We adopt the same canonical gauge as in the Wang-Yau quasi-local energy but allow the pulled back "killing vector" to the physical spacetime to be spacelike.
We define the new quasi-local energy along the Hamiltonian formulation of the Wang-Yau quasi-local energy. The new definition yields a positive definite surface energy density and a new divergence free current. Calculations for coordinate spheres in Kerr family spacetime are shown. In the spherical symmetric case, our definition reduces to a previous definition \cite{lundgren2007self}}. 

\section{Introduction}
In the path integral approach to the quantization of gravity, one must include spacetime metrics with black holes as well as metrics that can be continuously deformed to the flat metric. To avoid dealing with spacetime singularities that necessarily appear in a black hole metric, Gibbons and Hawking \cite{gibbonsHawking1977action} proposed to consider Euclidean action for gravity. An alternative approach may be employing the notion of quasi-local mass which is defined over a spacelike $2$-surface. In principle, the quasi-local mass of a $2$-surface $\Sigma$ encodes information about the gravitational field of a finite region bounded by $\Sigma$ which satisfies the dominant energy condition. In the case of Schwarzschild-like singularity, there clearly exists a family of topological $2$-spheres that foliates the spacetime outside the singularity. For more general types of singularity such as the ring singularity of Kerr, one can use surfaces of more general topology (e.g. torus for ring singularity) to foliate the spacetime outside the singularities.  One may then consider an infinite dimensional space of $2$-surfaces with its intrinsic and extrinsic geometric data to quantize gravity, instead of an infinite dimensional space of $4$-dimensional Lorentzian spacetimes.

The equivalence principle of Einstein gravity theory excludes the existence of a point-wise gravitational energy density. On the other hand, there exist well-defined notions of total gravitational energy for an asymptotically flat spacetime, e.g. ADM mass at spatial infinity and Bondi mass at null infinity. A quasi-local mass aims to measure gravitational field energy within a finite, extended region whose definition remains a major unresolved problem in general relativity \cite{Penrose1982unsolved,coley2017open}.
There exist several propositions for quasi-local energy/mass, e.g. Hawing mass\cite{hawking1968hawkingmass}, Bartnik mass\cite{bartnik1989newqlm}, Brown-York mass\cite{brown1993quasilocal}, Liu-Yau mass\cite{liuYau2003,liuYau2006} etc. However, they all possess some undesirable features \cite{Szabados:2009review}. For example, the Hawking mass could be negative even in the Minkowski spacetime; the Bartnik mass has no closed-form expression; the Brown-York mass depends on the choice of a spacelike $3$-manifold $\Omega$ bounded by the $2$-surface under consideration. Moreover, there exist surfaces in the Minkowski spacetime with strictly positive Liu--Yau mass as well as Brown--York mass \cite{murchadha2004comment}, violating the rigidity property that surfaces in the Minkowski spacetime should have zero mass. In order to overcome these difficulties,  Wang and Yau proposed a definition that incorporates both metric and momentum information about the $2$-surface under consideration \cite{wangyau2009cmp,wang2009quasilocalPRL}. Their definition is is intrinsic to the spacelike $2$-surface $\Sigma$ under consideration and most importantly, the positivity was proved.
Moreover, the Wang-Yau quasi-local mass was shown to be consistent with the ADM mass at spatial infinity \cite{WangYau:2010spatialinf} and Bondi mass at null infinity \cite{Chen:2010tz} and also exhibit reasonable small sphere limit at a limiting point \cite{ChenWangYau:2018smalllimit}.

However, the Wang-Yau quasi-local energy/mass as well as most other quasi-local energy/mass, is only defined for spacelike $2$-surfaces of spacelike mean curvature vector, including in particular future untrapped surfaces. To consider a moduli space of $2$-surfaces that encodes gravitational field information, one certainly needs to consider trapped surfaces or rather surfaces of timelike mean curvature.
This paper explores one possible extension of the original Wang-Yau definition to surfaces with timelike mean curvature vector. 
The rest of this paper is organized as follows. We briefly review the setup of Wang-Yau quasi-local mass in section \ref{sec:review}. We present our new quasi-local energy in section \ref{sec:new_inside_horizon}. The new definition is given in section \ref{subsec:new_defn}. A new mass surface density $\rho^{new}$ and a new divergence-free current $j^{new}$ are introduced in section \ref{subsec:rho_j}. Continuity cross the horizon between the definition here and the original Wang-Yau definition is discussed in section \ref{subsec:continuity_horizon}. Examples from well known black hole spacetime are shown in section \ref{subsec:example}. Lastly, we conclude and discuss in section \ref{sec:conclusion}.

\section{Review of Wang-Yau quasi-local mass}\label{sec:review}
The Wang-Yau quasi-local energy is defined as a difference between a reference term and a physical term. This ensures that for surfaces lying in the reference spacetime, typically taken as $\R^{3,1}$, the quasi-local energy vanishes.
Let us consider the reference spacetime first. Let $\Sigma$ be a spacelike $2$-surface from a general spacetime $\pst$ and let $X:\Sigma \to \R^{3,1}$ be an isometric embedding. Denote the image by $\Sigma_0=X(\Sigma)$. One also fixes a future-pointing, timelike, unit vector $T_0$ in the reference spacetime $\R^{3,1}$. Then the vector $T_0$ determines a unique timelike unit normal $u_0$ at each point of $\Sigma_0$ through
 $$T_0 = \sqrt{1+|\nabla\tau|^2} u_0 -\nabla \tau$$ 
where $\tau = -\langle X, T_0\rangle$ is the time function of $\Sigma_0$ along $T_0$. We denote covariant derivative on $\Sigma$ by $\nabla$ and covariant derivative in $\pst$ by $\overline{\nabla}$.
This leads to a \textit{canonical} basis $\{v_0,u_0\}$ for  the normal bundle $N\Sigma_0$ with
$$\langle v_0, u_0\rangle =\langle v_0, T_0 \rangle=0$$

Further, \cite{wang2009isometric} employed the frame independent mean curvature vector $H_0 = (\nabla_a e_a)^\perp$ and its conjugate vector
$J_0$ (reflection along the ingoing light cone in $N\Sigma_0$)
to define a \textit{reference} basis for the normal bundle, $$e_{H_0} = -\frac{H_0}{|H_0|}, e_{J_0} = \frac{J_0}{|H_0|}$$
where $|H_0|=|J_0|=\sqrt{|\langle H,H\rangle|}$.
For future-untrapped surface, 
$e_{H_0}$ is outward pointing while $e_{J_0}$ is future pointing. 
The basis transformation between the \textit{reference} basis and the \textit{canonical} basis is
\begin{align*}
    v_0 &= \cosh\theta_0 \, e_{H_0} - \sinh\theta_0 \, e_{J_0}\\
    u_0 &= -\sinh\theta_0 \, e_{H_0} + \cosh\theta_0 \, e_{J_0}
\end{align*}
with $$\sinh \theta_0 = \frac{-\Delta \tau}{|H_0|\sqrt{1+|\nabla \tau|^2}}$$

Now consider $\Sigma$ in the physical spacetime $\pst$ with spacelike mean curvature vector $H$. The canonical gauge condition 
\begin{equation}\label{eq:canonical_gauge}
 \langle T_0, H_0\rangle = \langle T,H\rangle   
\end{equation}
together with two other implicit conditions
\begin{align*}
     \langle T,T\rangle &=-1, \\
     T^T = T_0^T &= -\nabla\tau \quad (\text{projection to $T\Sigma$})
\end{align*}
uniquely determines a timelike unit vector $T$ at everypoint of $\Sigma$. The uniquely determined $T$ then picks a \textit{canonical} basis $\{u, v\}$ for $N\Sigma$ satisfying
\begin{align*}
    T &= \sqrt{1+|\nabla\tau|^2}u - \nabla \tau, \quad \\
     \langle v,u\rangle&=\langle v,T\rangle =0
\end{align*}
This solves the issue of choosing a frame for $2$-surface $\Sigma$ in $4$-dimensional spacetime $\pst$. 
Similarly, oen can define $\{e_H = -\frac{H}{|H|}, e_J=\frac{J}{|J|}\}$  as the \textit{reference} basis for $N\Sigma$. 
The basis transformation is
\begin{align}\label{eq:change_basis_spacelike}
    v &= \cosh\theta \, e_{H} - \sinh\theta \, e_{J} \nonumber \\
    u &= -\sinh\theta \, e_{H} + \cosh\theta \, e_{J}
\end{align}
with 
\begin{equation}\label{eq:angle_spacelike}
    \sinh \theta = \frac{-\Delta\tau}{|H|\sqrt{1+|\nabla \tau|^2}}
\end{equation}
determined by the canonical gauge \eqref{eq:canonical_gauge}.

By a theorem by Pogorelov \cite{pogorelov1964embedding}, a positive definite metric on a topological $2$-sphere $\Sigma$ with Gaussian curvature $K_\Sigma >-\kappa^2$ can be embedded into any hyperboloid $\eta_{\mu\nu} X^\mu X^\nu = C > - \frac{1}{\kappa^2}$ in $\R^{3,1}$. So an isometric embedding $X:\Sigma \to \R^{3,1}$ is far from unique. Wang and Yau defined the minimum of quasi-local energy among all possible isometric embeddings as the quasi-local mass. The variational problem can be solved through varying the time function $\tau=\langle X, T_0 \rangle$. The Euler-Lagrangian equation or the optimal embedding equation (OEE) is 
\begin{align}
    \nabla \cdot j &=0, \\
    j &= \rho\nabla\tau - \alpha_{v_0} + \alpha_v=\rho\nabla\tau - \alpha_{H_0} -\nabla\theta_0 + \alpha_H + \nabla\theta
\end{align}
where the connection 1-forms are defined as
\begin{align*}
    \alpha_{v_0} &= \langle \nabla^{\R^{3,1}} v_0, u_0\rangle = \alpha_{H_0} + \nabla \theta_0 \\
    \alpha_{v} &= \langle \overline{\nabla} v, u\rangle = \alpha_{H} + \nabla \theta\\
    \alpha_{H_0} &= \langle \nabla \frac{-H_0}{|H_0|}, \frac{J_0}{|J_0|}\rangle, \quad \alpha_H  = \langle \nabla \frac{-H}{|H|}, \frac{J}{|J|},  \rangle
\end{align*}

Lastly, we note that the canonical gauge condition \eqref{eq:canonical_gauge} is equivalent to
\begin{equation}\label{eq:canonical_gauge_u}
    \langle u, H \rangle = \langle u_0, H_0\rangle
\end{equation}
which matches the ``time'' component of mean curvature vectors. Difference in the other component measures the gravitational energy through the mass energy surface density $\rho$,
$$\rho = \frac{\langle H,v\rangle- \langle H_0,v_0\rangle}{\sqrt{1+|\nabla \tau|^2}}=\frac{\sqrt{|H_0|^2+\frac{\Delta\tau}{1+|\nabla\tau|^2}}-\sqrt{|H|^2+\frac{\Delta\tau}{1+|\nabla\tau|^2}}}{\sqrt{1+|\nabla\tau|^2}}$$
In terms of $\rho$ and $j$, the quasi-local energy is
$$\QLE = \frac{1}{8\pi}\int_\Sigma \rho + j\cdot \nabla\tau$$ 
which reduces to the quasi-local mass after imposing OEE $\nabla\cdot j=0$
$$\QLM = \frac{1}{8\pi}\int_\Sigma \rho$$ 

\section{Quasi-local mass for surfaces with time-like mean curvature vector}\label{sec:new_inside_horizon}
Here we present one possible extension of the Wang-Yau quasi-local mass to surfaces with time-like mean curvature vector, including in particular trapped surfaces inside an apparent horizon. Direct calculation of Hawking mass, Brown-York mass etc. for surfaces inside the horizon yields a blowup at the singularity \cite{gudapati2020quasi}. On the other hand, \cite{lundgren2007self} extended the Brown-York mass inside the horizon for spherical symmetric spacetimes based on physical considerations and yield finite results even at the singularity. The proposal here reduces to \cite{lundgren2007self} in spherical symmetric cases.

We first observe that the canonical gauge condition \eqref{eq:canonical_gauge} breaks down at an horizon where the mean curvature vector $H$ becomes null.
Consider a future-trapped $2$-surface $\Sigma$ with outgoing null expansion $\Theta_+<0$ and ingoing null expansion $\Theta_-<0$. A future-pointing timelike unit vector $T$ can be decomposed as
$$T = N u + \Vec{N}$$ 
where $N>0$, $\Vec{N}\in T\Sigma$ and $u$ is a future-pointing unit normal vector.  Let $v$ be the unit normal vector such that $\langle v,u\rangle = \langle v, T\rangle=0$. Then $\{u,v\}$ form an orthonormal basis for $N\Sigma$. In this basis, one has
$$H=-kv +pu, \quad \Theta_\pm=p \pm k$$ 
where $k=-\langle H, v\rangle$ and $p=-\langle H, u\rangle=\frac{\Theta_+ +\Theta_-}{2}<0$. Then
$$\langle H, T\rangle=\langle -kv+ pu,  N u + \Vec{N}\rangle = - Np >0$$
On the other hand,     $$\langle T_0,H_0\rangle  = -\Delta \tau$$
cannot be sign-definite over $\Sigma$.
Therefore, the canonical gauge \eqref{eq:canonical_gauge} fails for surfaces with timelike mean curvature vector.


Then one obvious rescue is to allow $T$ to change signature, in particular, one can set $T$ to be everywhere spacelike when $H$ is everywhere timelike. 
In other words, we regard $T_0$ as the killing vector that generates the conserved energy in $\R^{3,1}$ and we determine the pulled back ``killing'' vector $T$ through imposing \eqref{eq:canonical_gauge}. This is similar to that the Brill-Lindquist coordinate $\partial_t$ turns to spacelike at the ergosphere.

\subsection{Proposed definition for surfaces with time-like mean curvature}\label{subsec:new_defn}
As argued above, for surfaces with timelike mean curvature vector $H$, we instead take $\langle T, T\rangle =1$. Note that in the \textit{reference} basis for $N\Sigma$, $e_H=\frac{-H}{|H|}$ is now future-pointing timelike while $e_J=\frac{J}{|H|}$ is now outward-pointing spacelike. The notation is that $|H|=|J|=\sqrt{|\langle H,H\rangle|}$.
The vector $T$ now decomposes as 
\begin{align}
    T 
    &= \sqrt{1-|\Vec{N}|^2}\,v + \Vec{N}, \quad 1-|\Vec{N}|^2>0
\end{align}
The basis transformation then becomes
\begin{align}\label{eq:change_basis}
    v &= \cosh\theta \, e_{J} - \sinh\theta \, e_{H} \nonumber \\
    u &= -\sinh\theta \, e_{J} + \cosh\theta \, e_{H}
\end{align}
with 
\begin{equation}\label{eq:angle_timelike}
    \sinh \theta = \frac{\Delta \tau}{|H|\sqrt{1-|\Vec{N}|^2}}
\end{equation}
determined by the canonical gauge \eqref{eq:canonical_gauge}.

We now work out the quasi-local energy along the Hamiltonian argument.
As reviewed in \cite{zhao2024some}, the quasi-local energy can be defined through the generalized mean curvature vectors 
\begin{align*}
    \Xi &= k u + K(v,\cdot)-(trK) v \in T\pst \\
    \Xi_0 &= k_0 u_0 + K_0(v_0,\cdot)-(trK_0)\, v_0 \in \R^{3,1}
\end{align*}
where
\begin{align*}
    \quad k_0=-\langle H_0, v_0\rangle, \quad & K_0(v_0,\cdot) = \langle \nabla^{\R^{3,1}}_{v_0} u_0,\cdot\rangle \\
    k=-\langle H, v\rangle, \quad  & K(v,\cdot) = \langle \overline{\nabla}_v u,\cdot\rangle 
\end{align*}
That is,
\begin{align*}
     8\pi \, QLE = \int_\Sigma -\langle \Xi_0, T_0\rangle - \int -\langle \Xi, T \rangle & \\
    = \int -\sqrt{1+|\nabla \tau|^2}\,\langle v_0, H_0\rangle + \alpha_{v_0}(-\nabla \tau) & - \int -\sqrt{1-|\Vec{N}|^2}\,\langle u, H\rangle + \alpha_v (\Vec{N})\\
    = \int \sqrt{1+|\nabla\tau|^2} \cosh \theta_0 |H_0| + \theta_0\,\Delta \tau  -\alpha_{\hat{e}_3}(\nabla \tau) & - \int -\sqrt{1-|\Vec{N}|^2}\cosh\theta|H| - \theta\, \nabla\cdot \Vec{N} + \alpha_{e_3}(\Vec{N})
\end{align*}
If we extremize the integrand of the second integral, as in \cite{wang2009isometric}, one reaches at
$$\sinh\theta = \frac{-\Delta\cdot \Vec{N}}{|H|\sqrt{1-|\Vec{N}|^2}}$$
which agrees with \eqref{eq:angle_timelike} when $\Vec{N}=-\nabla \tau$. From now on, we will make the choice of $\Vec{N} = -\nabla\tau$. Use the following inequality proved in \cite{wang2009isometric}
$$\int_{\hat{\Sigma}} \hat{k} = \int_\Sigma \sqrt{1+|\nabla\tau|^2} \cosh \theta_0 |H_0| + \theta_0\,\Delta \tau  -\alpha_{\hat{e}_3}(\nabla \tau)$$
One reaches the new quasi-local energy definition as follows.
\begin{defn}[Quasi-local energy for $2$-surface with time-like mean curvature]
    \begin{align}\label{eq:QLE_timelike}
     QLE^{new} & = \frac{1}{8\pi} \int_{\hat{\Sigma}} \hat{k} - \frac{1}{8\pi} \int_\Sigma -\sqrt{1-|\nabla\tau|^2}\langle H, u\rangle - \alpha_v (\nabla\tau)  \nonumber \\
    &=\frac{1}{8\pi} \int_{\hat{\Sigma}} \hat{k} - \frac{1}{8\pi}\int -\sqrt{1-|\nabla\tau|^2}\cosh \theta |H| -\nabla\theta\cdot \nabla\tau - \alpha_{J}(\nabla\tau)
\end{align}
where 
\begin{equation}
    \sinh \theta = \frac{\Delta \tau}{|H|\sqrt{1-|\nabla \tau|^2}}
\end{equation}
and 
\begin{equation}
    \alpha_J(\cdot) = \langle \overline{\nabla}_{(\cdot)} \frac{J}{|H|}, \frac{-H}{|H|}\rangle
\end{equation}
\end{defn}


\subsection{Surface mass density and current}\label{subsec:rho_j}
Similar to the Wang-Yau quasi local energy, one would like to look for a critical point among various embeddings $X$  or different killing vector $T_0$. The first variation of the new definition with respect to the time function $\tau=-\langle X, T_0\rangle$ again yields a divergence free current 
\begin{equation}
    \nabla\cdot j^{new} =0
\end{equation}
where 
\begin{align*}
    j^{new} &= \frac{\nabla \tau}{\sqrt{1+|\nabla \tau|^2}}|H_0|\cosh\theta_0 -\frac{\nabla\tau}{\sqrt{1-|\nabla\tau|^2}}|H|\cosh\theta -\nabla(\theta_0-\theta) - \alpha_{H_0}+\alpha_J\\
    &=\Tilde{\rho}_1 \nabla \tau + \nabla \sinh^{-1} \frac{\Tilde{\rho}_2 \,\Delta\tau}{|H||H_0|} -\alpha_{H_0}+\alpha_J
\end{align*}
with
$$\Tilde{\rho}_1 = \frac{\sqrt{|H_0|^2+\frac{(\Delta\tau)^2}{1+|\nabla\tau|^2}}}{\sqrt{1+|\nabla\tau|^2}}-\frac{\sqrt{|H|^2+\frac{(\Delta\tau)^2}{1-|\nabla\tau|^2}}}{\sqrt{1-|\nabla\tau|^2}}$$
$$\Tilde{\rho}_2 = \frac{\sqrt{|H_0|^2+\frac{(\Delta\tau)^2}{1+|\nabla\tau|^2}}}{\sqrt{1-|\nabla\tau|^2}}+\frac{\sqrt{|H|^2+\frac{(\Delta\tau)^2}{1-|\nabla\tau|^2}}}{\sqrt{1+|\nabla\tau|^2}}$$

Further define a new mass surface density
\begin{align}
    \rho^{new} &= \frac{|H_0|\cosh\theta_0}{\sqrt{1+|\nabla\tau|^2}} + \frac{|H|\cosh\theta}{\sqrt{1-|\nabla\tau|^2}}\\
    &=  \frac{\sqrt{|H_0|^2+\frac{(\Delta\tau)^2}{1+|\nabla\tau|^2}}}{\sqrt{1+|\nabla\tau|^2}}+\frac{\sqrt{|H|^2+\frac{(\Delta\tau)^2}{1-|\nabla\tau|^2}}}{\sqrt{1-|\nabla\tau|^2}}
\end{align}
Then the new quasi-local energy can be written as
\begin{equation}
    QLE^{new} =  \frac{1}{8\pi} \int_\Sigma \rho^{new} + j^{new} \cdot \nabla \tau
\end{equation}
which reduces to $$QLM^{new} = \frac{1}{8\pi} \int_\Sigma \rho^{new}$$
if the Euler-Lagrangian equation $\nabla \cdot j^{new}=0$ is imposed. 
Note that the new mass surface density is nonnegative. The positivity of new quasi-local \textit{mass} follows from this. We recall that the surface mass density $\rho$ in Wang-Yau quasi-local mass is not positive definite. However, the new quasi-local energy with non-critical $\tau$ is not positive definite.
Calculation with Schwarzschild coordinate spheres indicate that the critical value of $\tau$ seems to corresponds to a local maximum in QLE$^{new}$. We will study the critical point in a separate work. 

\subsection{Continuity across an horizon}\label{subsec:continuity_horizon}
Having defined a quasi-local energy/mass for both trapped and untrapped surfaces, one would naturally ask whether the two definition is continuous cross the horizon, where the mean curvature vector is null.
A careful analysis of the Wang-Yau quasi-local energy in the limit of $|H|\to 0^+$ is shown in \cite{ZAY:limit_horizon}. 
The analysis there shows that the Wang-Yau quasi-local energy blows up to $+\infty$ unless the horizon can be isometrically embedded into $\R^3$.

A similar conclusion applies to the new definition presented here. One can rewrite the new quasi-local energy as in \cite{ZAY:limit_horizon}:
\begin{align}
    & 8\pi \, QLE^{new} = \nonumber \\
    & \int_{\hat{\Sigma}} \hat{k} + \int_\Sigma \sqrt{(\Delta\tau)^2 + |H|^2(1-|\nabla\tau|^2)} + \nabla\tau \cdot \nabla\ln|\Theta_-| +\frac{1}{2}\langle \overline{\nabla}_{\nabla\tau}\, l^-,l^+\rangle +\tau\Delta y \label{eq:QLE_trouble_term}
\end{align}
where $y=\ln \bigg[\frac{-\Delta\tau}{\sqrt{1-|\nabla\tau|^2}} + \sqrt{|H|^2+\frac{(\Delta\tau)^2}{1-|\nabla\tau|^2}}\bigg]$. 
Meanwhile the new Eular-Lagrangian equation can be rewritten as 
\begin{equation}
    \Delta y = \nabla\cdot \big( \Tilde{\rho}_1 \nabla\tau -\alpha_{H_0}-\nabla\theta_0 \big)
\end{equation}
A similar analysis would yield that 
$\int \tau \Delta y  \to -\infty$
and hence  $QLE^{new} \to  -\infty $
unless $\tau$ approaches to a constant function at the horiozn.
We summarize in the following theorem.
\begin{theorem}\label{thm:limit_horizon_spacelike}
Assume that an isometric embedding of $\Sigma$ remains smooth such that $|\nabla\tau|$, $\Delta\tau$ all remain bounded as $\Sigma$ approaches an horizon from both sides, i.e. $|H|\to 0$. Further assume that $|\Theta_-|=-\Theta_-$ is bounded away from zero for the horizon under consideration. If the horizon cannot be embeded into $\R^3$, then 
$$QLE\to +\infty, \textbf{ as } |H|\to 0^+ $$
while 
$$QLE^{new}\to -\infty, \textbf{ as } |H|\to 0^- $$
and moreover, both Euler-Lagrangian equations
$$\nabla \cdot j=0, \quad \nabla\cdot j^{new}=0$$
are not solvable near the horiozn.
\end{theorem}
\begin{rmk}
    If instead, the horizon can be isometrically embedded into $\R^3$ and moreover,     $$x:=\lim_{|H|\to 0}\,\frac{\Delta\tau}{|H|\sqrt{1+|\nabla\tau|^2}} < \infty$$ i.e. $\tau$ approaches to a constant function in a faster or comparable rate as $|H|\to 0$, then
$$ lim_{|H|\to 0} QLE = \frac{1}{8\pi} \int_{{\Sigma}} |H_0| $$
\end{rmk}
Therefore, unless the horizon can be isometrically embedded into $R^3$, the Wang-Yau quasi-local energy blows up while the new quasi-local energy blows down as $\Sigma$ approaches the horizon from outside or inside and hence there is no continuity cross the horizon.

\subsection{Examples of black hole spacetime}\label{subsec:example}
\begin{figure}
    \centering
    \includegraphics[width=0.8\textwidth]{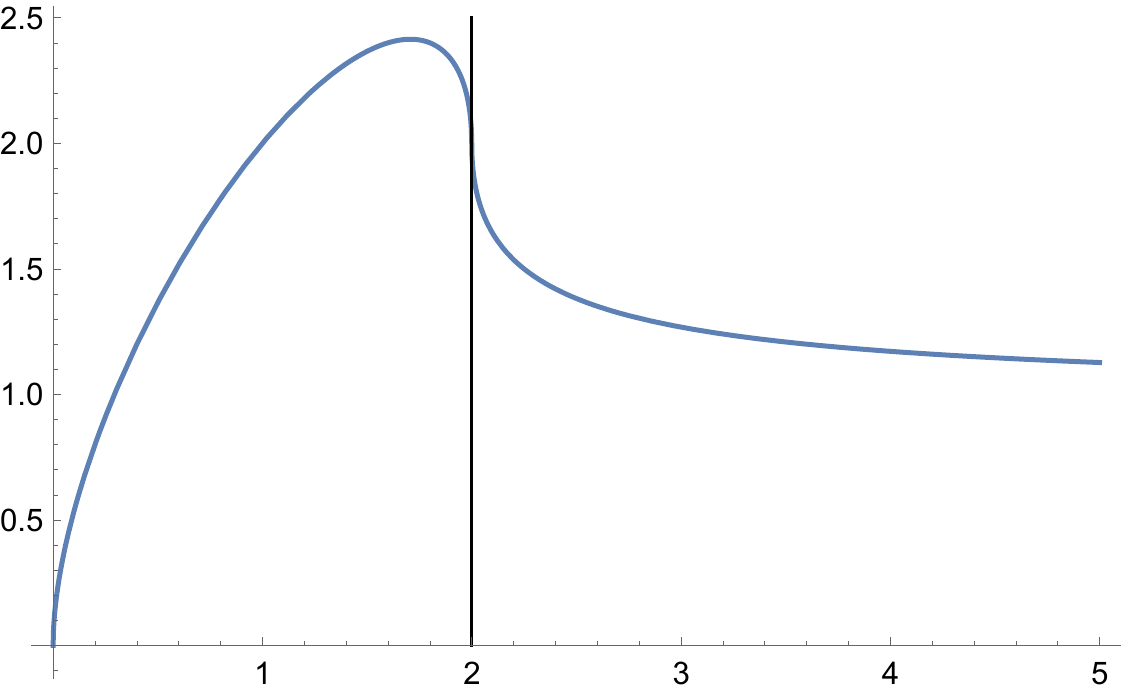}
    \caption{Quasi-local energy with $\tau=0$ for coordinate spheres in the Schwarzschild spacetime. The vertical black line indicates where the horizon locates.}
    \label{fig:Schwarzschild}
\end{figure}

\begin{figure}
    \centering
    \includegraphics[width=0.8\textwidth]{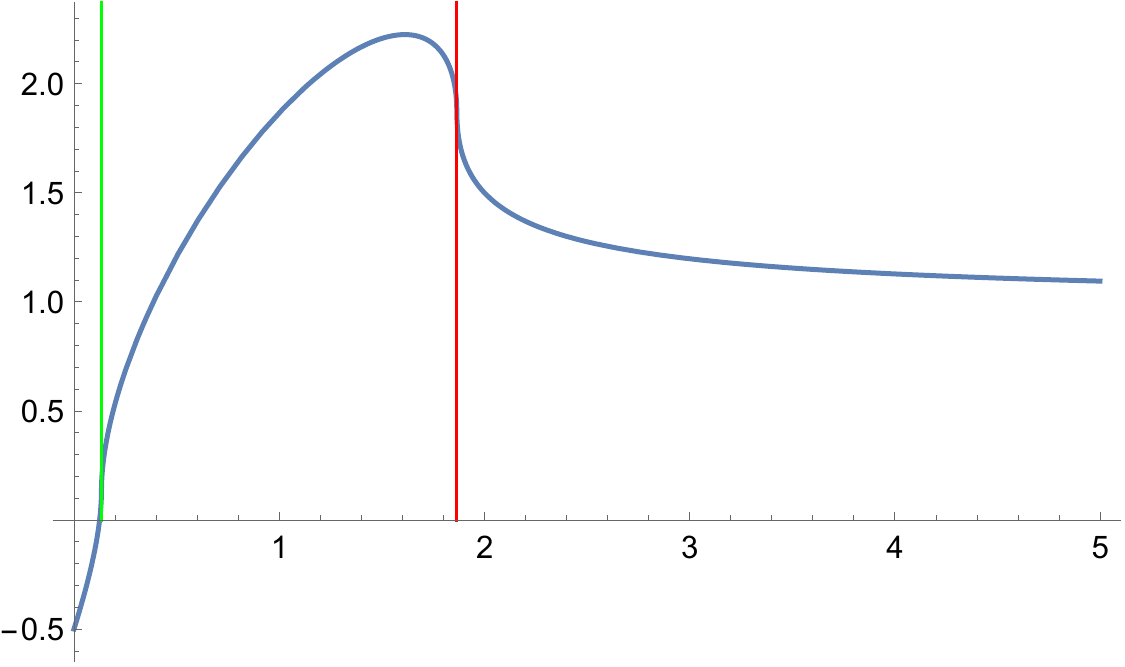}
    \caption{Quasi-local energy with $\tau=0$ for coordinate spheres 
 in the Reissner-Nordst\"{o}rm with $Q=M/2$. The red and green vertical line indicates the outer and inner horizon. Note that the mean curvature vector turns back to spacelike inside the inner horizon. The negative value at $r=0$ is $-|Q|$. }
    \label{fig:RN}
\end{figure}

\begin{figure}
    \centering
    \includegraphics[width=0.8\textwidth]{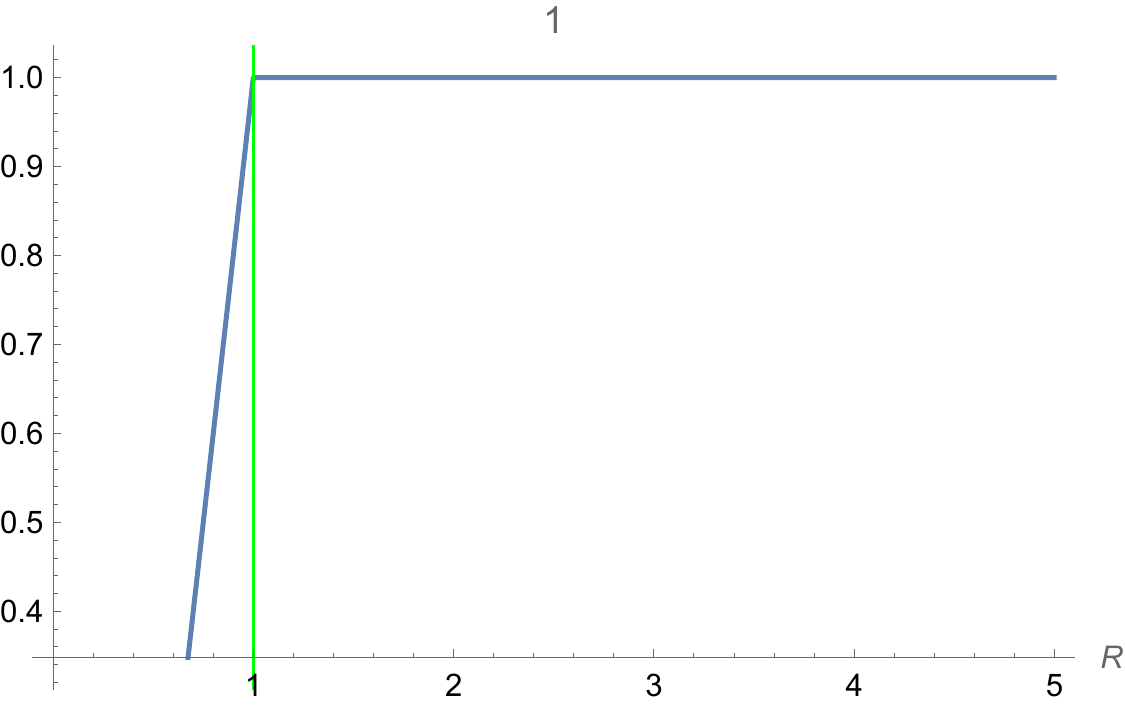}
    \caption{Quasi-local energy with $\tau=0$ for coordinate spheres 
 in the Extremal Reissner-Nordst\"{o}rm with $Q=M$. The negative value at $r=0$ is $-|Q|$. }
    \label{fig:extremeRN}
\end{figure}

\begin{figure}
    \centering
    \includegraphics[width=0.8\textwidth]{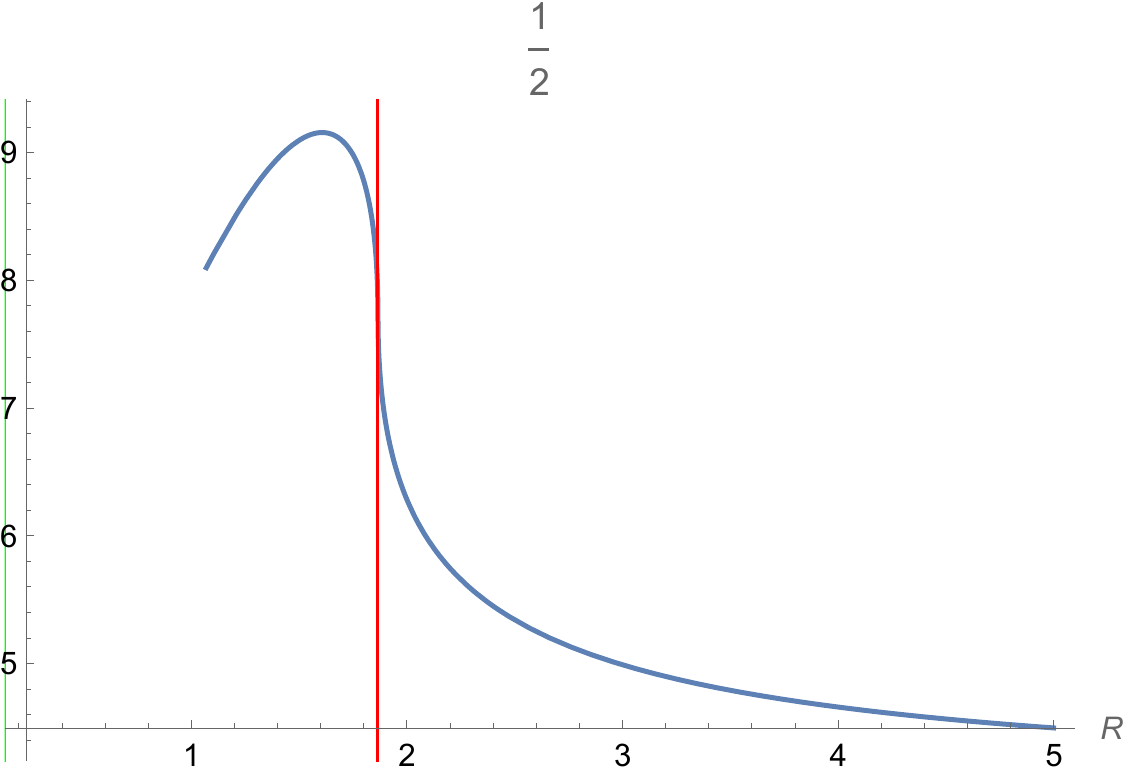}
    \caption{Quasi-local energy with $\tau=0$ for coordinate spheres 
 in the Kerr spacetime with $a=M/2$. Calculation in done numerically with Mathematica. Ingoing Kruskal coordinates is used. The calculation stops when $\tau=\const$ yields complex values of QLE, indicating $\Sigma_r$ can not be embedded into $\R^3$. Calculation for Kerr-Newmann black hole with $Q\neq 0$ is similar.}
    \label{fig:kerr}
\end{figure}

Recall that for coordinate spheres in an stationary and axisymmetric spacetime, i.e. a Kerr family black hole, constant $\tau$ solves the optimal embedding equation, both inside and outside the horizon since $\nabla\cdot \alpha_{H}=-\nabla\cdot \alpha_{J} \equiv 0$ while
$\theta=\pm \sinh^{-1}\frac{\Delta\tau}{|H|\sqrt{1+|\nabla\tau|^2}}=0$ and $\alpha_{v_0}= \langle \nabla^{\R^{3,1}}v_0, T_0\rangle =0$ when $\tau$ is constant and $|H|\neq 0$. 
Here we show calculations for coordinate spheres from slowly rotating Kerr family for which one can simply take constant $\tau$. The quasi-local energy with constant $\tau$ reduces to
\[QLM= \begin{cases} 
      \frac{1}{8\pi} \int |H_0| + |H|, & \text{ inside the horizon}\\
      \frac{1}{8\pi} \int |H_0| - |H|, & \text{ outside the horizon}
   \end{cases}\] 
We show in Figure \ref{fig:Schwarzschild}-\ref{fig:kerr} calculations for coordinate spheres of Schwarzschild blackhole, Reissner-Nordst\"{o}rm black hole with $Q=M/2$, extreme Reissner-Nordst\"{o}rm black hole with $Q=M$ and Kerr black hole with $a=M/2$, respectively.
Note that in these examples, the horizon can be embedded into $\R^3$ and hence the QLE attains a finite limit as $|H|\to 0$
$$\lim_{|H|\to 0} \,\QLE = \frac{1}{8\pi}\int_\Sigma |H_0|$$
The infinite slope at the horizon only manifests the fact that the horizon for these black hole spacetime happen to be minimal surfaces, whose first variation of area vanishes. Since the $x-axis$ is the area radius $r$, one has $\delta r\propto \delta |\Sigma|=0$ and hence $\frac{\delta \QLM}{\delta r}\to \infty$ at the horizon.

For Schwarzschild black hole of mass $M$,
\[QLM= \begin{cases} 
      r[1-\sqrt{1-\frac{2M}{r}}], & r \geq 2M \\
      r[1+\sqrt{\frac{2M}{r}-1}], & r \leq 2M 
   \end{cases}\] 
with the maximum at $r=(1+\frac{1}{\sqrt{2}})M$.
For Reissner-Nordst\"{o}rm of charge $Q$ and mass $M$,
\[QLM= \begin{cases} 
      r[1-\sqrt{1-\frac{2M}{r}+\frac{Q^2}{r^2}}] & r \geq r_+ \\
      r[1+\sqrt{-1+\frac{2M}{r}-\frac{Q^2}{r^2}}] & r_-\leq r \geq r_+\\
      r[1-\sqrt{1-\frac{2M}{r}+\frac{Q^2}{r^2}}] & r \leq r_- \end{cases}\]
with the zero at $r_0 = \frac{Q^2}{2M}< r_- = M-\sqrt{M^2-Q^2}$. So
the negative mass for Reissner-Nordst\"{o}rm happens inside the inner horizon, where mean curvature vector of $\Sigma$ is spacelike. However, this does not contradict previous proofs on positivity of Brown-York or Wang-Yau mass because the singularity is essentially ``naked'' to surfaces inside the outer horizon, violating assumptions in those proofs. The radius where quasi-local energy turns negative is also when
the textbook mass function $$M-\frac{Q^2}{2r}$$ turns negative. The value at the singularity is $-|Q|$. These two spherical symmetric examples recover the results of \cite{lundgren2007self} and reveal interesting features that worths further exploration.
For the Kerr black hole, when rotation $a<\sqrt{3}/2 M$, the horizon has positive Gaussian curvature, so we can still take $\tau=0$ for surfaces inside but near the horizon. Further inside, the calculation breaks down since those coordinate spheres cannot be embedded into $\R^3$, i.e. $\tau=0$ is not a valid solution to the Euler-Lagrangian equations.

\section{Concluding remarks}\label{sec:conclusion}
As revealed in calculations for the Kerr family black hole, the new definition of quasi-local energy exhibits some interesting features that may worth further investigation. For example, the new quasi-local mass seems to attain the maximum for a $2$-surface $\Sigma$ in the trapped region. Also, the calculation for Reissner-Nordst\"{o}rm black hole raises the question whether the Wang-Yau quasi-local energy for $\Sigma$ with spacelike mean curvature vector is bounded from below even when the spacetime singularity is not shielded from  $\Sigma$. 

The definition here is based on the setup of Wang-Yau quasi-local energy and hence reveals a similar pattern as one takes the limit of $\Sigma$ approaching to an horizon. The basic assumption here is that the canonical gauge condition of Wang and Yau \cite{wang2009isometric} still holds formally. It certainly needs further study to investigate what the optimal gauge condition to impose for trapped surfaces in order to pick a orthonormal basis $\{u,v\}$ for the normal bundle. This study is a first attempt in trying to define a quasi-local energy for surfaces inside the horizon.

We can now formulate the alternative approach to the path integral quantization of gravity using the concept of quasi-local mass. Consider an infinite dimensional space of topological $2$-spheres with geometric data $(\sigma,\alpha,|H|)$ where $\sigma$ is a positive-definite metric, $\alpha$ is a one-form and $|H|$ is a scalar function on $\Sigma$. The proposal is to rewrite the formal path integral over spacetime metrics $g$ as a formal path integral over geometric data of topological $2$-sphere $\Sigma$:
\begin{align*}
    \int D[g] e^{i I[g]} = \int D[\sigma] D[\alpha] D[|H|] D[\tau] D[T_0] e^{i\QLM}
\end{align*}
In principle, one may consider another integral over the topology of $\Sigma$. We note that this is different string theory, which quantizes the world sheet geometry instead of the spacetime geometry. 

\clearpage

\clearpage
\clearpage

\clearpage


\bibliographystyle{alpha}
\bibliography{reference}
\end{document}